\documentclass[aps,prd,preprint,superscriptaddress,showpacs,floatfix,nobibnotes,nofootinbib]{revtex4-2}
\usepackage{geometry}
\usepackage{latexsym}
\usepackage{amsmath}
\usepackage{amssymb}
\usepackage{graphicx}
\usepackage{setspace}
\usepackage{slashed}
\usepackage{ulem}
\usepackage{float}
\usepackage{inputenc}
\usepackage{caption}
\usepackage{subcaption}
\usepackage{bm}
\usepackage{epsfig}
\usepackage{array,etoolbox}
\preto\tabular{\setcounter{magicrownumbers}{0}}
\newcounter{magicrownumbers}

\newcommand{\nn}{\nonumber}
\newcommand{\bea}{\begin{eqnarray}}
\newcommand{\eea}{\end{eqnarray}}
\newcommand{\bi}{\begin{itemize}}
\newcommand{\ei}{\end{itemize}}

\usepackage[svgnames]{xcolor}

\begin{document}

\title{Effects of different 3D QED vertex ansaetze on critical coupling}

\author{M.E. Carrington}
\email[]{carrington@brandonu.ca} 
\affiliation{Department of Physics, Brandon University, Brandon, Manitoba, R7A 6A9 Canada}
\affiliation{Department of Physics \& Astronomy, University of Manitoba, Winnipeg, Manitoba, R3T 2N2 Canada}
\affiliation{Winnipeg Institute for Theoretical Physics, Winnipeg, Manitoba}
 
\author{A.R. Frey}
\email[]{a.frey@uwinnipeg.ca} \affiliation{Department of Physics, University of Winnipeg, Winnipeg, Manitoba, R3B 2E9 Canada}
\affiliation{Department of Physics \& Astronomy, University of Manitoba, Winnipeg, Manitoba, R3T 2N2 Canada}
\affiliation{Winnipeg Institute for Theoretical Physics, Winnipeg, Manitoba}

\author{B.A. Meggison}
\email[]{brett.meggison@gmail.com} 
\affiliation{Department of Physics \& Astronomy, University of Manitoba, Winnipeg, Manitoba, R3T 2N2 Canada}
\affiliation{Winnipeg Institute for Theoretical Physics, Winnipeg, Manitoba}

\date{\today}

\begin{abstract}

We study the semi-metal/insulator phase transition in graphene using a Schwinger-Dyson approach. 
We consider various forms of vertex ansaetze to truncate the hierarchy of Schwinger-Dyson equations. 
We define a Ball-Chiu-type vertex that truncates the equations without violating gauge invariance. 
We show that there is a family of these vertices, parametrized by a continuous parameter that we call $a$, all of which satisfy the Ward identity.
We have calculated the critical coupling of the phase transition using different values of $a$. 
We have also tested a common approximation in which only the first term in the Ball-Chiu ansatz is included. 
This vertex is independent of $a$, and, although it is not gauge invariant, it has been used many times in the literature because of the numerical simplifications it provides. 
We have found that, with a one-loop photon polarization tensor, the results obtained for the critical coupling from the truncated vertex and the full vertex with $a=1$ agree very well, but other values of $a$ give significantly different results. 
We have also done a fully self-consistent calculation, in which the photons are backcoupled to the fermion degrees of freedom, for one choice $a=1$. Our results show that when photon dynamics are correctly taken into account, it is no longer true that the truncated vertex and the full Ball-Chiu vertex with $a=1$ agree well. 
The conclusion is that traditional vertex truncations do not really make sense in a system that does not respect Lorentz invariance, like graphene, and the need to include vertex contributions self-consistently is likely inescapable. 

\end{abstract}


\normalsize
\maketitle

\normalsize

\section{Introduction}
\label{section-introduction}

Graphene has been studied by many physicists since its discovery, in part because of its technological applications, and also because of its interest to theorists. 
The lattice structure of graphene produces a fermion dispersion relation that is linear near the lowest energies, called Dirac points. 
The low-energy dynamics is  well described by a continuum quantum field theory called reduced QED in which the electronic quasiparticles have a linear dispersion relation and are restricted
to move in the two-dimensional plane of the graphene sheet, while the photons are
free to move in three dimensions \cite{marino, miransky}.

The coupling constant in the effective theory is dimensionless and non-perturbatively large, which is part of the reason that graphene is interesting to theorists. 
Several different methods are available to study non-perturbative systems. 
In this paper we use Schwinger-Dyson (SD) equations, one of the most commonly used continuum non-perturbative approaches, and study some issues that arise when the method is applied to study graphene. 

For a given theory, the SD equations provide a coupled infinite hierarchy of integral equations for non-perturbative dressing functions that describe the modification of the bare theory by the interactions. 
The set of equations has to be truncated in a way that respects gauge invariance, and includes the essential physics that is relevant to the problem at hand. 
The truncated equations involve a coupled set of dressing functions, 
each of which has support on a three or four-dimensional momentum phase space. 
In the case of graphene, the number of dressing functions is larger than in standard QED because the condition $v_F\ne c$ breaks Lorentz invariance. The number of dressing functions increases further in an anisotropic situation, where the $x$ and $y$ eigenvalues of the Fermi velocity tensor differ.

SD equations cannot be solved analytically and numerical solutions are notoriously difficult to obtain. 
Different approximations can be introduced to truncate the SD hierarchy, and to reduce  the number of dressing functions that must be included. Another common approach is to neglect the frequency dependence of some or all dressing functions.
In this paper we are particularly interested in the choice of the vertex ansatz that is used to truncate the SD equations. 
Some of the early calculations that used bare vertices include \cite{Khveshchenko_2004,liu_2009,gamayun_2009,Wang_2011,Khveshchenko_2009,gamayun_2010}. The authors of Ref. \cite{CuevasFiniteT} also used bare vertices but included finite temperature effects. In Refs. \cite{gonz_2010,gonz_2012} non-perturbative effects were including using a resummation of ladder graph contributions to vertices, instead of a SD approach. 

The idea to construct an ansatz for a non-perturbative vertex in terms of the fermion dressing functions was introduced 40 years ago \cite{ball-chiu-1,ball-chiu-2}. The Ball-Chiu (BC) vertex is constructed to truncate the SD equations in a covariant gauge theory, without violating gauge invariance. A simplification that is commonly employed is to take only the first term in the vertex (see Eq. (\ref{BC-MEC1})) which is the easiest to handle numerically. This ansatz, which is usually called the BC$_1$ vertex approximation, can be modified in a straightforward way for anisotropic theories like graphene, and was used in Refs. \cite{Wang_2012,mec1,mec2}. 
There are also components of the vertex that are transverse (give zero when contracted with the momentum of the photon line) and cannot be determined from gauge invariance. Various modifications of the BC vertex that include transverse contributions have been used in QED \cite{pennington-1,pennington-2,pennington-3}. 
An ansatz that includes transverse vertex components was recently used in a SD calculation in covariant reduced QED  \cite{Albino_2022}. 
We also mention that in Ref. \cite{BjornDisorder} it was argued that for disordered semi-metals the vertex function in the SD equation for the fermion self-energy can be replaced with a Gaussian function called a disorder correlator.

In this paper we study modifications of the BC vertex, that go beyond the BC$_1$ approximation, and give gauge invariant truncations of a non-Lorentz invariant theory like graphene. 
We show that for the effective theory we use to describe graphene there is no unique way to construct a BC-type vertex using only the constraint that the Ward identity must be satisfied.  
There is a family of ansaetze, parametrized by a continuous parameter that we call $a$, all of which satisfy the Ward identity. 

We also consider the effect of an anisotropic Fermi velocity. The calculation was formulated in Ref. \cite{xiao1} where the simple approximation of bare vertices was used. An alternate formulation that also uses bare vertices can be found in Ref. \cite{Sharma_2017}. 
In Refs. \cite{MC-AF-BM-anioG,brett2} we discussed how the structure of the SD equations  must be modified when anisotropy is present, even at the level of the BC$_1$ approximation.

To evaluate the effect of the parameter $a$ on a physical result we focus on the critical coupling at which graphene goes through a quantum phase transition from a semi-metal state to an insulating state.
The critical coupling is defined as the value of the coupling for which the dressed electrons acquire a non-zero mass. The value of this coupling is of  interest because the ability to produce an insulating state would have enormous significance in the development of graphene based electronic devices.
Our calculations show that the value of the critical coupling for the phase transition depends fairly strongly on the choice of the parameter $a$. 
This result indicates that a correct formulation of the calculation will likely require self-consistently determined vertex functions.

We emphasize that the value of the critical coupling produced
by any calculation based on an effective theory is not expected to be exact, since
there are potentially important screening effects that are necessarily ignored. 
The point of this work is to explore the validity of the SD approach in the context of graphene and determine directions for future work. We further note that the insulating state has not been seen experimentally, in spite of the fact that many theoretical calculations predict that the critical coupling should be physically realizable. It is interesting however that a small band gap has been observed at finite temperature in bilayer graphene \cite{Tenasini_2022}.

We work in Euclidian space throughout this paper. We use the notation $P_\mu = (p_0,\vec p)$, $P^2 = p_0^2+p^2$, and similarly for other momentum variables. We define $Q=K-P$ and use the shorthand $dK = dk_0 dk_1 dk_2/(2\pi)^3$. 
We use natural units: $\hbar=c=1$.

\section{Notation}
\label{notation-sec}

We consider mono-layer graphene at zero temperature and zero chemical potential. 
Close to the critical point, the system can be described using a low energy effective theory with broken Lorentz invariance.  
The photons move in three dimensions, while the electrons are restricted to the two dimensional plane of the graphene sheet and have non-relativistic velocities.
The bare Feynman rules (in covariant gauge) are
\bea
\label{bareFR}
&& S^{(0)}(P) = -\big[i\gamma_\mu M_{\mu\nu} P_\nu\big]^{-1}\,\nn\\[2mm]
&& G^{(0)}_{\mu\nu}(P)  = \big[\delta_{\mu\nu}-\frac{P_\mu P_\nu}{P^2}(1-\chi)\big]\,\frac{1}{2\sqrt{P^2}}\,\nn \\[1mm]
\label{barevert}
&& \Gamma^{(0)}_\mu = M_{\mu\nu}\gamma_\nu
\eea
where $M = (1,v_F,v_F)_{\rm diag}$.
The photon propagator for reduced QED is obtained by integrating out momentum modes perpendicular to the plane of the graphene sheet \cite{marino,miransky}. 
It corresponds to a nonlocal action because the photon propagates outside the plane of the graphene sheet.
Since $v_F\ll 1$ it is a good approximation to use the Coulomb approximation, which means to take the static limit of the bare photon propagator. In our numerical calculations we make this approximation [see Sec. \ref{setup2}, and in particular equation (\ref{G00-cou})].
The inverse dressed propagators are defined as
\bea
&& S^{-1} = S^{(0)\;-1} +\Sigma \\
&& G^{-1} = G^{(0)\,-1} + \Pi_{\mu\nu}
\eea
where $\Sigma$ and $\Pi_{\mu\nu}$ are the fermion self-energy and photon polarization tensor, respectively. 

These dressing functions can be decomposed using a set of projection operators, each of which is multiplied by a different scalar function. 
The  theory that describes graphene involves more dressing functions than standard QED, because the (non-unity) Fermi velocity breaks Lorentz invariance. 
In standard QED the fermion self-energy can be written $\Sigma = -i (A-1) \slashed{P} + D$ where the functions $A$ and $D$ are momentum dependent dressing functions that reduce to $A=1$ and $D=0$ in a bare massless theory. The photon polarization tensor can be written as the product of a four-dimensionally transverse projection operator and one dressing function, as $\Pi_{\mu\nu} = (\delta_{\mu\nu} P^2 - P_\mu P_\nu) \bar \Pi$.
To describe graphene, these expressions must be modified. 
For electrons in graphene, the self-energy is defined as  
\bea
\Sigma = -i \gamma_\mu M_{\mu\alpha}(F_{\alpha\nu} - \mathbb{I}_{\alpha\nu})P_\nu + D
\label{sigma-def}
\eea
where $F = (Z,A,A)_{\rm diag}$. 
It is easy to see that in the limit $v_F=1$ and $Z=A$, Eq. (\ref{sigma-def}) reduces to the conventional covariant expression. 
The photon polarization tensor in graphene is written in terms of two dressing functions, which correspond to coefficients of two projection operators. These two projectors are transverse and longitudinal with respect to the momentum three-vector, while both are four-dimensionally transverse as required by gauge invariance. 
The three components of the fermion self-energy (Eq. (\ref{sigma-def})) satisfy integral equations that are calculated by taking the appropriate projections of its SD equation. In the limit $v_F\ll 1$ the contribution of the three-dimensionally transverse component of the photon propagator to all three components of $\Sigma$ is suppressed by a factor $v_F^2$, and the only component of the photon propagator that is needed to calculate the fermion dressing functions is
\bea
&&   G_{00} = \delta_{\mu 0}\delta_{\nu 0} G_{\mu\nu} = \frac{p^2}{P^2}\frac{1}{2\sqrt{P^2} + \frac{P^2}{p^2} \Pi_{00}}\,.
\label{G00}
\eea
This  means that only one component of the polarization tensor needs to be calculated, which leaves a total of four dressing functions $(Z,A,D,\Pi_{00}$). 
%

The effect of an anisotropic Fermi velocity has been considered by several authors. 
The motivation is that anisotropy could reduce the critical coupling and make the insulating state easier to realize. 
One introduces an additional fermion dressing function, so that the function $A$ gets replaced by two functions $A_1$ and $A_2$. 
The definition of the fermion self-energy is obtained from (\ref{sigma-def}) using $M=(1,v_1,v_2)_{\rm diag}$ and a non-diagonal definition of the matrix $F$ of the form
\bea
F &=& \left[
\begin{array}{ccc}
	Z & 0 & 0 \\
	0 & A_1 & A_2 \\
	0 & -A_2 & A_1 \\
\end{array}
\right]\,.
\eea
We define the Fermi velocity $v_F = \sqrt{v_1 v_2}$, the anisotropy parameter $\eta=v_1/v_2$, and the coupling $\alpha = e^2/(4\pi v_F)$.
The naive definition $F=(Z,A_1,A_2)_{\rm diag}$ does not reproduce the isotropic result in the limit $v_1=v_2=v_F$; furthermore, the naive diagonal definition renormalizes the anisotropy, while the correct form renormalizes the Fermi velocity (and principle axes) but not the anisotropy \cite{MC-AF-BM-anioG}.

The conclusion to date is that anisotropy likely increases the critical coupling \cite{xiao1,MC-AF-BM-anioG,brett2}.  
However, the most sophisticated calculations that have so far been done rely on a vertex ansatz that is not fully gauge invariant (see discussion below). 
One of the goals of this paper is to determine how reliable this kind of vertex ansatz is.

\section{Schwinger-Dyson equations}
\label{setup2}

The SD integral equations for the fermion self-energy and photon polarization tensor are
\bea
\label{fermion-SD}
\Sigma(p_0,\vec p) &=& 	e^2\int dK \,G_{\mu\nu}(q_0,\vec q)\,M_{\mu\tau}\,\gamma_\tau \, 	
	S(k_0,\vec k) \,\Gamma_\nu\\
	\label{photon-SD}
\Pi_{\mu\nu}(p_0,\vec p) &=& -e^2\int dK \,{\rm Tr}\,\big[S(q_0,\vec q) \, M_{\mu\tau} \, 
	\gamma_\tau \, S(k_0,\vec k)\, \Gamma_\nu\big]
\eea
where $\Gamma_\mu$ is a three-point function that will be discussed below. 
As discussed in Sec. \ref{notation-sec}, in the limit $v_F\ll 1$ we need only the zero-zero component of the photon propagator and photon polarization tensor. We discuss below two further approximations that can be applied to the photon propagator.
\begin{enumerate}
\item The Coulomb approximation means that the frequency dependence of the bare propagator is dropped. This is a very common approximation, especially in the condensed matter community, and is motivated by the idea that the photon velocity is much greater than the fermion velocity. The zero-zero component of the photon propagator in Eq. (\ref{G00}) becomes
\bea
G^{\rm coulomb}_{00} = \frac{1}{2\sqrt{p^2} + \Pi_{00}}\,.
\label{G00-cou}
\eea

\item The one-loop approximation means that the function $\Pi_{00}$ is replaced by the result of a one-loop calculation where the integrand is constructed with bare lines and vertices. This is also a common approximation and is motivated by the vanishing fermion density of states at the Dirac points. The one-loop result for the zero-zero component of the polarization tensor is
\bea
\Pi^{{\rm one loop}}_{00}(p_0,\vec p) &=& \frac{\pi \alpha}{v_F} \frac{v_F^2 p^2}{\sqrt{P_\mu M_{\mu 
	\rho} M_{\rho \nu} P_\nu}}\,. \label{1loopPi}
\eea
\end{enumerate}

Next we discuss the vertex function $\Gamma_\mu$ that appears in Eqs. (\ref{fermion-SD} and \ref{photon-SD}). 
The full hierarchy of SD equations  provides an integral equation for this vertex function in terms of a four-point function. 
As discussed in Sec. \ref{section-introduction}, the coupled nature of the equations requires that some truncation is introduced. 
The equation for the three-vertex is extremely difficult to solve, and therefore we will use an ansatz that expresses this vertex in terms of the fermion dressing functions, so that the two equations (\ref{fermion-SD} and \ref{photon-SD}) form a closed set. 
The BC vertex \cite{ball-chiu-1,ball-chiu-2} is constructed to satisfy the Ward identity, which in our notation is written
$-iQ_\mu \Gamma_\mu(P,K) = S^{-1}(P) - S^{-1}(K)$,
and truncates the SD equations without violating gauge invariance. 
The form of the $BC$ vertex must be modified when the underlying theory is not Lorentz invariant. 
The authors of Ref. \cite{mec1} used the ansatz
\bea
&& \Gamma_\mu(P,K) = \frac{1}{2}\big(F(p_0,\vec p)_{\mu\alpha}^T+
	F(k_0,\vec k)_{\mu\alpha}^T\big)M_{\alpha\beta}\gamma_\beta \label{BC-MEC1} \\
	\label{ballchiu}
	&& ~~ +\bigg[\frac{1}{2}(P+K)_\alpha\big[F(p_0,\vec p)_{\alpha\beta}^T-F(k_0,\vec k)_{\alpha\beta}^T\big]M_{\beta\rho} \gamma_\rho + i(D_p-D_k)\bigg]\frac{(P+K)_\mu}{P^2-K^2} \nn
\eea
motivated by the idea that this is the simplest extension of the BC vertex that satisfies the Ward identity. 
However, it is easy to see that the ansatz in (\ref{BC-MEC1}) will satisfy the Ward identity if the factor multiplying the square bracket is changed as
\bea
\frac{(P+K)_\mu}{P^2-K^2} \to \frac{\tilde M(a)_{\mu \lambda}\tilde M(a)^{\lambda\nu}(P+K)_\nu}
{(P \tilde M  \tilde M  P - 
K  \tilde M  \tilde M  K)}
\label{BC-mod}
\eea
where $\tilde M(a)_{\mu\nu} = (1,a , a )_{\rm diag}$ and $a$ is any real number, and we have used the shorthand notation $P  \tilde M  \tilde M  P = P_\mu\tilde M(a)_{\mu\lambda}\tilde M(a)_{\lambda\nu} P_\nu$ (and likewise for $K$). 
The original ansatz in (\ref{BC-MEC1}) corresponds to 
$\tilde M(1) = \mathbb{I}$. A more natural choice might be $\tilde M(v_F) = M$. 
Any value of $a$ is equally valid from the point of view of gauge invariance. 

In many SD calculations the term in square brackets in (\ref{BC-MEC1}) is dropped.
This is usually called the BC$_1$ approximation for the vertex:
\bea
 \Gamma^{\rm Short}_\mu(P,K) &=& \frac{1}{2}\big[F(p_0,\vec p)_{\mu\alpha}^T+
	F(k_0,\vec k)_{\mu\alpha}^T\big]M_{\alpha\beta}\gamma_\beta  \,.
	\label{shortballchiu}	
\eea
This expression does not satisfy the Ward identity, but is numerically much easier to work with. The point is that the denominator $1/(P^2-K^2) = 1/(p_0^2+p^2-k_0^2-k^2)$ has singularities along a curve in the domain of the integral over $(k_0,k)$ for which $P^2-K^2=0$. These singularities are integrable, but they are  difficult to deal with numerically. 
We can also see that the square bracket in (\ref{BC-MEC1}) will go to zero at $(p_0=k_0,p=k)$, and perhaps other places along the curve defined by the equation $P^2=K^2$. 
The non-perturbative fermion dressing functions do not have simple symmetry properties that would allow us to identify the curves along which the square bracket in the  BC vertex is zero. 
We note that in a Lorentz invariant calculation, where the fermion dressing functions depended on only one variable, the cancellation of the zeros in the numerator and denominator occurs at one easily identifiable point in the domain of the integral. 
The integrals obtained using a non-covariant BC vertex are numerically much more difficult to calculate, and we further expect that some choices of the value of $a$ could be numerically more stable than others. 
We point out that the authors of Ref. \cite{mec1} (using the value $a=1$) found that dropping the second term in (\ref{BC-MEC1}) had virtually no effect on the critical coupling that was obtained from the calculation.  
It therefore seems possible that the issue described above could have no practical significance. 
In fact it turns out that results can be strongly dependent on the value of the parameter $a$.

The full set of SD equations are calculated by substituting Eqs. (\ref{G00-cou}, \ref{1loopPi}, \ref{BC-MEC1}, \ref{BC-mod}) into (\ref{fermion-SD} and \ref{photon-SD}), multiplying by the appropriate projection operators, and contracting over all indices. The procedure is straightforward but very lengthy. All calculations were done using FORM \cite{FORM}. 
The results, for $a=1$, are given in Appendix \ref{SDequations-sec}.

\section{Results}
\label{results}

\begin{figure}[H]
	\begin{center}
		\includegraphics[width=0.75\linewidth]{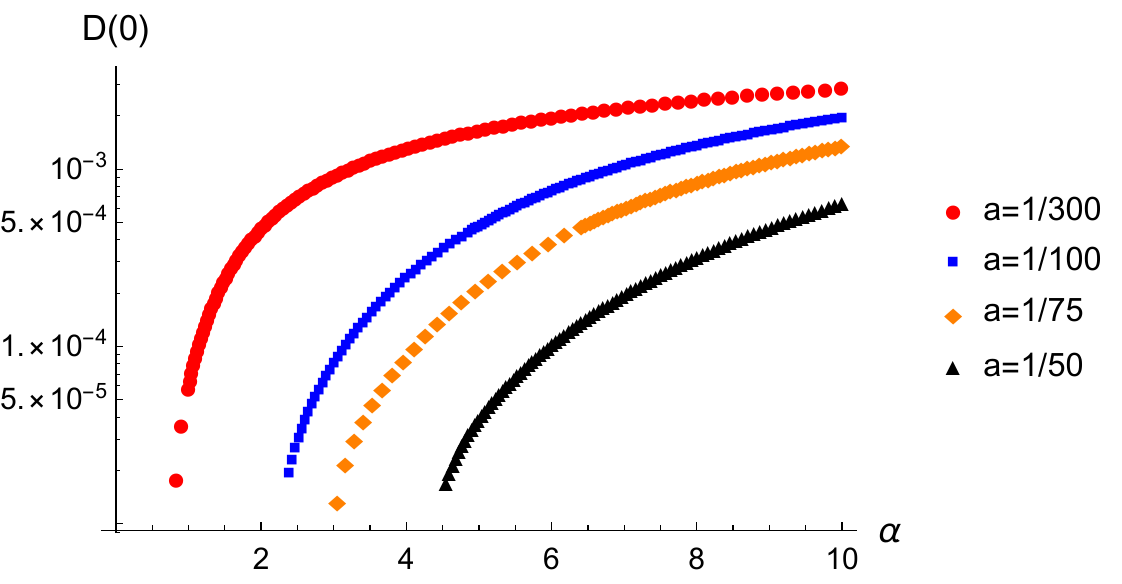}
	\end{center}
	\caption{The condensate as a function of the coupling, from a calculation including only one dressing function.}
	\label{Donly}
\end{figure}

To explore the dependence of the critical coupling on the value of $a$, we begin by adopting a very simple approximation in which all fermion dressing functions other than $D$ are set to their bare values $Z=A=1$. We also use the Coulomb approximation and the one-loop photon polarization tensor instead of self-consistently solving the photon SD. These approximations are known to give unrealistically large values for the critical coupling. 
The point of these calculations is only to test the effect of the value of $a$ in the simplest possible way. 
Figure \ref{Donly} shows that reducing $a$ reduces the critical coupling, and the dependence is fairly strong. 
In Table \ref{table1} we give the critical coupling for each value of $a$. The second number in the bracket indicates the error for each result. The numerical method we use to find a value of the critical coupling and its error is explained in Appendix \ref{numerics-sec}.
\begin{table}[H]
\begin{center}
\begin{tabular}{|c|c|c|c|c|}
\hline 
~~ $a$ ~~ & ~~~~ 1/300 ~~~~  &  ~~~~ 1/100  ~~~~ & ~~~~ 1/75  ~~~~ &  ~~~~ 1/50  ~~~~ \\
 \hline
$\alpha_c$  & (0.74, 0.09) & (2.22, 
  0.17) & (2.94, 0.11) & (4.41, 0.14) \\
 \hline 
 \end{tabular}
\end{center}
\caption{The critical couplings for different values of $a$, obtained from the results shown in Fig. \ref{Donly}.\label{table1}}
\end{table}

\begin{figure}[H]
	\begin{center}
		\includegraphics[width=0.75\linewidth]{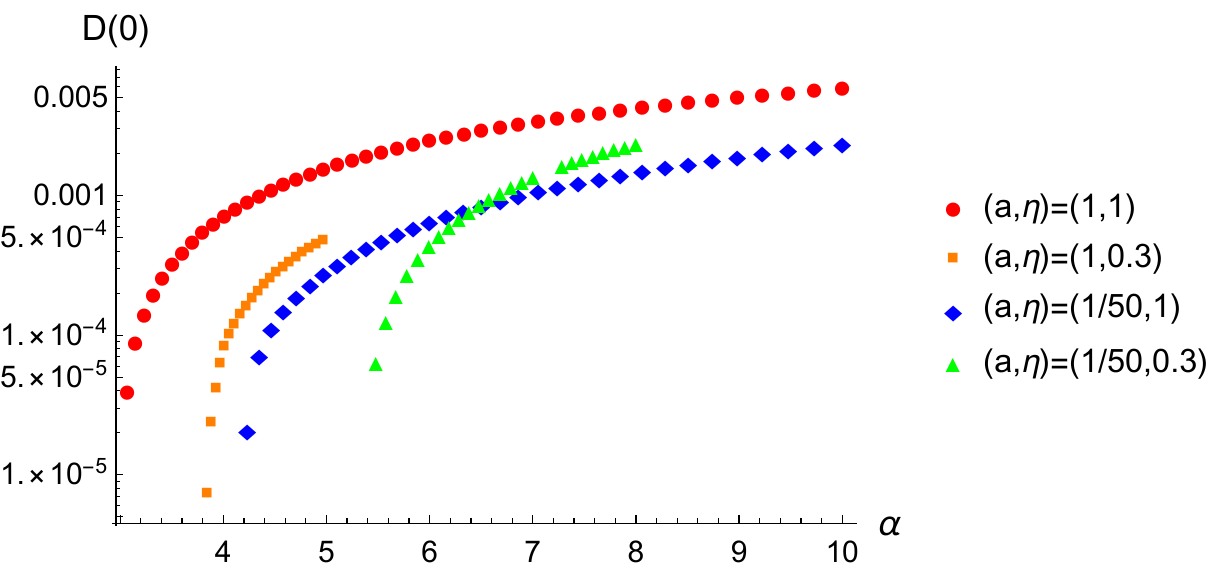}
	\end{center}
	\caption{The condensate as a function of the coupling using the vertex in Eqs. (\ref{BC-MEC1} and \ref{BC-mod}). The red (circles) show $(a,\eta)=(1,1)$, orange (squares) are $(a,\eta) = (1,0.3)$, blue (diamonds) are $(a,\eta)=(1/50,1)$, and green (triangles) are $(a,\eta)=(1/50,0.3)$.}
	\label{G00coul}
\end{figure}

Next we solve the coupled set of equations for all three fermion dressing functions but again using the Coulomb approximation and the one-loop approximation for the photon self-energy. 
In Fig. \ref{G00coul} we show the results for four choices of the parameters $(a,\eta)$. The figure shows once again that the form of the vertex ansatz has a significant effect on the critical coupling, 
in fact, larger than the shift in the critical coupling obtained by introducing anisotropy. This is most clearly demonstrated by comparing sets of curves with the same value of $a$ and different values of $\eta$, which are (red, orange) and (blue, green), and curves with different $a$ but the same $\eta$ (red, blue) and (orange, green). 
The results show clearly that a change in anisotropy has less effect on the critical coupling than a change in the value of $a$.
It is also interesting that the direction of the $a$ dependence is the opposite of what is seen in Fig. \ref{Donly} - reducing $a$ increases the critical coupling. This is not unexpected since the calculation shown in Fig. \ref{Donly} does not include Fermi velocity renormalization, which is known to have a significant effect on the critical coupling. 
We also comment that values of $a$ smaller than $1/50$ do not converge without using a huge number of lattice points (see Sec. \ref{setup2} for a discussion of this point). The corresponding results for the critical coupling are shown in Table \ref{table2}. The effect of increasing the anisotropy is slightly greater for the smaller value of $a$. 
\begin{table}[H]
\begin{center}
\begin{tabular}{|c|c|c|c|c|}
\hline 
~~ $(a,\eta)$ ~~ & ~~~~ (1,1) ~~~~ & ~~~~ (1,0.3) ~~~~  &  ~~~~ (1/50,1)    ~~~~  &  ~~~~ (1/50,0.3)    ~~~~ \\
 \hline
$\alpha_c$ & (2.999, 0.068) & (3.814, 0.073) & (4.204, 0.029) & (5.361, 0.117) \\
 \hline 
 \end{tabular}
\end{center}
\caption{The critical couplings for different values of $(a,\eta)$ from the results shown in Fig. \ref{G00coul}. \label{table2}}
\end{table}

\begin{figure}[H]
	\begin{center}
		\includegraphics[width=0.75\linewidth]{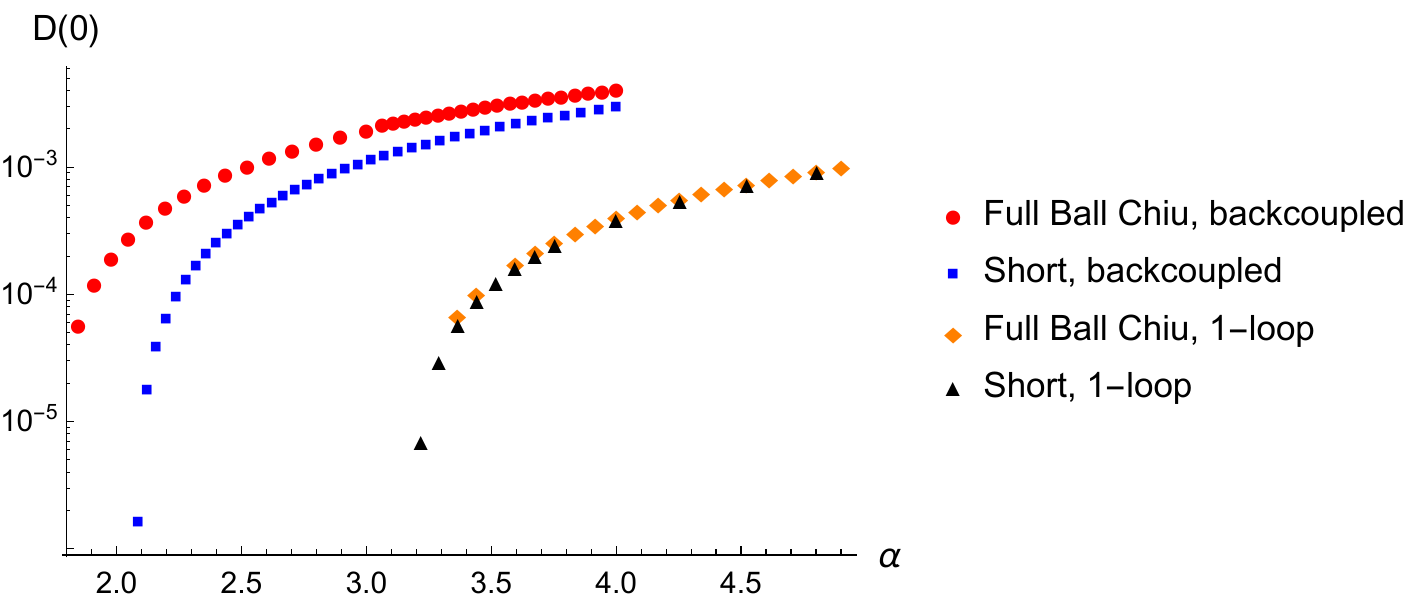}
	\end{center}
	\caption{The condensate versus the coupling using the BC$_1$ vertex, and the full BC vertex with $a=1$, with and without backcoupling. }
	\label{figisoFBCFBC}
\end{figure}

Another important thing to check is the effect of using the one-loop approximation for the polarization tensor.
The polarization tensor should be determined self-consistently from its integral equation, which is given in the last equation of Eq. (\ref{FBCIsoEqns}). 
In \cite{mec2} it was found, using the BC$_1$ vertex in equation (\ref{shortballchiu}), that the effect of using a self-consistently determined photon polarization tensor reduced the coupling fairly dramatically (from 3.19 to 1.99). Figure \ref{figisoFBCFBC} shows $D(0)$ versus $\alpha$ for the approximate BC$_1$ vertex, and the full BC vertex with $a=1$, in both cases with and without the one-loop approximation for the polarization tensor.  The figure shows that when the photon degrees of freedom are backcoupled, the result from the approximate vertex does not agree well with the full BC vertex. The corresponding results for the critical coupling are shown in Table \ref{table3}.
\begin{table}[H]
\begin{center}
\begin{tabular}{|c|c|c|c|c|}
\hline 
~~  ~~ & ~~~~ BC backcoupled ~~~~ & ~~~~ BC$_1$ backcoupled ~~~~  &  ~~~~ BC one loop    ~~~~ & ~~~~ BC$_1$ one loop ~~~~ \\
 \hline
$\alpha_c$ & (1.782, 0.065) & (2.085, 0.037) & (3.19, 
  0.19) & (3.199, 0.092) \\
 \hline 
 \end{tabular}
\end{center}
\caption{The critical couplings for the data shown in figure \ref{figisoFBCFBC}. \label{table3}}
\end{table}

\section{Conclusions}
\label{conclusions}

We have studied the semimetal-insulator phase transition in graphene by solving a set of coupled Schwinger-Dyson equations. The full set of equations is numerically prohibitively difficult to solve, and for this reason many different approximations are common in the literature. We have studied the effect of some of these approximations, with particular emphasis on possible choices of the  ansatz for the three-point vertex function, which is introduced to truncate the hierarchy of SD equations. We have shown that there is no unambiguous way to extend the definition of the Ball-Chiu vertex for a theory in which Lorentz invariance is broken, like graphene. There is a family of vertices, parametrized by a continuous variable (that we call $a$), all of which satisfy the Ward identity. 
Our calculations show that the value of the critical coupling depends strongly on the value of  $a$, and the effect of anisotropy also depends quantitatively on $a$. 
The conclusion is that to predict the critical coupling, one must include a true non-perturbative three-vertex. 
This could be done either by truncating the SD equations by introducing some ansatz for the four-point function, or using a three-particle reducible (3PI) effective action approach. We note that one advantage of working with the 3PI effective action is that all truncations occur at the level of the action, which means that gauge invariance is automatically preserved to the level of the truncation \cite{Smit2003,Zaraket2004}. 
While it is true that the renormalization of NPI actions is a notoriously difficult problem \cite{4pi-renorm1,4pi-renorm2}, 
the issue becomes largely trivial for a theory defined in less than four dimensions, like graphene. 
A 3PI effective action approach might be a promising method to study phase transitions in graphene 
when starting from a continuum field theory. 
We also mention that if one is only interested in finding the critical coupling at which the condensate goes to zero, then bifurcation theory could be used (see Ref. \cite{PhysRevD.94.125009} for an application of bifurcation theory to three dimensional QED).

\appendix

\section{The SD equations}
\label{SDequations-sec}
The SD equations for an anisotropic theory, obtained with a BC vertex ansatz are given in equation (\ref{FBCIsoEqns}). 
These expressions were calculated from equations (\ref{fermion-SD}, \ref{photon-SD}) by defining the appropriate projection operators and contracting over Lorentz and Dirac indices. All calculations were done using FORM \cite{FORM}.
The denominator of the fermion propagator is written
\bea
S_p &=& p_0^2 Z_p^2 + v_1^2 \left(p_1 A_{1}+p_2 A_{2}\right){}^2 
+ v_2^2 \left(p_2 A_{\text{1p}} - p_1 A_{\text{2p}}\right){}^2+D_p^2 \,.\label{SpAniso}  
\eea
The subscript $D$, for each dressing function, indicates a difference function defined as, for example
\bea
Z_D = \frac{Z_P - Z_K}{P_\mu \tilde M(a)_{\mu \rho} M\tilde(a)_{\rho \nu} P_\nu-K_\mu \tilde M(a)_{\mu \rho} \tilde M(a)_{\rho \nu} K_\nu}\,
\label{Z-diff}
\eea
and similarly for all fermion dressing functions. In the integral equation for $\Pi_{00}$ we used the notation $Z_{QD}$ to indicate the function in equation (\ref{Z-diff}) with $P$ replaced by $Q$, and similarly for the other fermion dressing functions. The subscript  $S$ represents the sum of two dressing functions, for example $Z_S = Z_P + Z_K$, $Z_{SQ} = Z_Q + Z_K$, etc.
We note that all dependence on the parameter $a$ appears in equation (\ref{Z-diff}), and all dependence on the anisotropy parameter is in equation (\ref{SpAniso}). 
The parameter $\chi$ should be set to zero to obtain the equations that correspond to the truncated vertex in equation (\ref{shortballchiu}), and one otherwise. The isotropic limit corresponds to $\eta=1$ and $A_2=0$. 

\bea
Z(P) &=& 1+2 \alpha \pi v_F\int \frac{dK}{S_k}\frac{q^4 \left(k_0 Z_K \left((k_0+p_0)^2 \chi  Z_D+Z_S\right)+2 D_D \chi  D_K \left(k_0+p_0\right)\right)}{2 p_0
	\left(Q^2\right)^{3/2} \left(2 q^2+\sqrt{Q^2} \Pi ^{00}_Q\right)}\nn \\
A_1(P) &=& 1-2 \alpha \pi v_F\int \frac{dK}{S_k}\frac{q^4}{2 p^2 \left(Q^2\right)^{3/2} \left(2 q^2+\sqrt{Q^2} \Pi^{00}_Q\right)} \Big(A_{2 K} \left(k_2 p_1-k_1 p_2\right) \left((k_0+p_0)^2 \chi  Z_D+Z_S\right) \nn \\
&&+A_{1K} \vec{k} \cdot \vec{p} \left((k_0+p_0)^2 \chi  Z_D+Z_S\right)+ \nn \\
	&&k_0 (-\chi ) \left(k_0+p_0\right) Z_K \left(A_{2 D} \left(k_2 p_1-k_1
	p_2\right)+A_{1D} \left(\vec{k} \cdot \vec{p} + p^2\right)\right)\Big)\nn \\
A_2(P) &=& 2 \alpha \pi v_F\int \frac{dK}{S_k}\frac{q^4}{2 p^2 \left(Q^2\right)^{3/2} \left(2 q^2+\sqrt{Q^2} \Pi ^{00}_Q\right)} \Big(k_0 (-\chi ) \left(k_0+p_0\right) Z_K \big(A_{2 D} \left(k \cdot p+p^2\right)\nn \\ 
&&-\left(k_2 p_1-k_1 p_2\right) A_{1 D}\big)+A_{2 K}
	k \cdot p \left(\chi  Z_D \left(k_0+p_0\right){}^2+Z_S\right)\nn \\
	&&-\left(k_2 p_1-k_1 p_2\right) A_{1 K} \big(\chi  Z_D
	\left(k_0+p_0\right){}^2 +Z_S\big)\Big)\nn \\
D(P) &=&  2 \alpha \pi v_F\int \frac{dK}{S_k}\frac{q^4 \left(D_K \left(\chi  Z_D \left(k_0+p_0\right){}^2+Z_S\right)-2 D_D k_0 \chi  \left(k_0+p_0\right) Z_K\right)}{4 q^2 \left(Q^2\right)^{3/2}+2
	Q^4 \Pi ^{00}_Q}\nn \\
\Pi_{00}(P) &=& -16 \alpha \pi v_F\int \frac{dK}{S_k S_q}2 \chi  D_K D_Q \left(k_0+q_0\right){}^2 Z_{DQ}-2 k_0 q_0 \chi  \left(k_0+q_0\right){}^2 Z_K Z_Q Z_{DQ}- \nn \\
&&4 k_0 \chi  D_Q D_{DQ} \left(k_0+q_0\right)
Z_K-4 q_0 \chi  D_K D_{DQ} \left(k_0+q_0\right) Z_Q+2 D_K D_Q Z_{SQ}-2 k_0 q_0 Z_K Z_Q Z_{SQ}. \nn\\
\label{FBCIsoEqns}
\eea

\section{Numerics}
\label{numerics-sec}
In this section we give a brief summary of the main aspects of our numerical procedure. 

All integrations were done using a Gauss-Legendre method. 
We used spherical coordinates, so that there are two  independent variables $(p_0,|\vec p|)$ and three integration variables $(k_0,|\vec k|, x=\hat p\cdot\hat k$). 
The $k_0$ and $k$ integrations were done on a logarithmic scale to increase sensitivity to the small momentum part of the phase space where the dressing functions change most. 
Interpolations were done using a combination of linear interpolation, in regions where the function to be interpolated was fairly flat, and Pade approximates, in regions the function changed more sharply. 
Convergence was obtained with $(N_{p_0}=30,N_{p}=30)$ points in the external momentum space and $(N_{k_0}=100,N_{k}=100,N_x=36)$ internal grid points. Adequate computational speed was achieved by parallelizing the code using MPI (Message Passing Interface). 

To calculate the critical coupling from a given set of data that gives the value of the condensate $D(0)$ for different values of the coupling $\alpha$, we use the following procedure. 
We invert the array to obtain a numerical representation of $\alpha[D(0)]$, construct an interpolated function, and extrapolate to find the critical coupling $\alpha_c\equiv \alpha[0]$. In all cases, the data that we need to interpolate are very smooth, and different interpolation methods give results for the critical coupling that agree to very high precision. 

It is clear, however, that the accuracy of the result for the critical coupling does not depend on the accuracy of the interpolated function. For example, if the last three points on the blue curve in figure \ref{G00coul} were missing, the extrapolated critical alpha would be much too small. 
Numerically however, the last points are the most difficult to calculate because of the phenomena called critical slowing down, which occurs when the solution is very close to the trivial solution for which the dressing function $D(p)$ is zero over its full domain. 
One way to quantify the error in the extrapolated result for the critical coupling would be to remove the last calculated point, and compare with the previous result. If the data stopped at a point where the curvature of the data was large, this would give a significantly different critical coupling. However, if the last few points in the data
give a line that is fairly straight but not close to vertical, this procedure will indicate a small error even though the extrapolated critical alpha will not be very accurate. 
An alternative estimate is the absolute difference between the extrapolated result and the smallest value of $\alpha$ in the data set. The error calculated this way is related to the inverse slope of the data at small $\alpha$;  it will be small if the curve drops steeply to the horizontal axis, and large if the curve is fairly flat. 

In all of the calculations we have done, the absolute difference of the extrapolated value and the smallest calculated value is larger than the error obtained by removing the last point, and all of the errors given in the results section are absolute difference errors. 
We also note that in this calculation the error is always positive, because the critical coupling produced by the extrapolation can only be smaller than the true critical coupling.

\begin{acknowledgments}
This work has been supported by the Natural Sciences and
Engineering Research Council of Canada Discovery Grant program 
 from grants 2017-00028 and 2020-00054. 
This research was enabled in part by support provided by WestGrid
(www.westgrid.ca) and the Digital Research Alliance of Canada (alliancecan.ca).
\end{acknowledgments}

\bibliography{anisotropic}

\end{document}